\documentclass{hcij}
\usepackage{microtype}

\begin{document}

\articletitle{Exploring the Future of Remote \\User Research}

\authorlist{Nikolas Martelaro}

\affiliationlist{Human-Computer Interaction Institute, Carnegie Mellon University}


\articleabstract{
    The rapid move to remote work due to COVID-19 social distancing policies has slowed or stopped most in-person qualitative user research activities. 
    The limitation on in-person activities has pushed many user researchers to consider how they can continue their research remotely and presents an opportunity for user research teams to (re)design their research processes for remote work. 
    In this position statement, I outline prior work in remote qualitative user research methods and discuss how this work can help us to (re)design new tools, methods, and frameworks that can enable remote work. 
    I then propose a set of research questions that can further our knowledge and abilities to conduct effective remote user research. 
    With creative development, we can create remote research methods that afford new ways of understanding user experience today and build more environmentally conscious and accessible research practices for a future where remote user research is much more common.
}

\titlenotes
\titlebackground{This article is based on a workshop paper presented at the Microsoft Research New Future of Work 2020 Symposium and the author's Doctoral thesis.}

\noindent\titleacknowledgments{I would like to thank the many user researchers I had conversations with during the last few months and the 56 members of the professional user research community who answered the survey exploring the impact of COVID-19 and social distancing on their work.
I want to thank Judy Chun and Chi Huang for there help in exploring these ideas and helping conduct interviews and the survey. 
I would also like to thank Sarah Fox, Patrick Carrington, and Wendy Ju for conversations that helped me form the ideas in this position statement.}

\newpage

\thickhrrule

\tableofcontents

\thickhrrule

\articlebodystart


\section{Introduction}
\label{sec:introduction}

The COVID-19 pandemic and subsequent social distancing practices have pushed information and technology workers to remote work practices.
These workers include user researchers who conduct interviews and observations with potential users to generate a holistic understanding of peoples everyday experiences and inform the development of new product and service designs \cite{beyer_contextual_1997, kuniavsky_observing_2003, laurel_design_2003}.
Much of a user research team's work is conducted in-person through either lab-based of field-based interactions with users.
Due to COVID-19, many of these in-person research activities have been significantly reduced or even stopped.
But with the rapid move to remote work, we may ask ourselves how important is it that user research work is done in-person?
Over the years, researchers and practitioners have explored how to conduct user research remotely without needing to be physically co-located with users.
Today, many user researchers who work on desktop, web, and mobile software products can readily do their work using a selection of tools and technologies such as high-quality video chat, screen sharing, and remote data telemetry \cite{black_remote_2017}.
However, the same cannot be said of user researchers who work on products and services that fundamentally require understanding users in physical contexts such as the home, the workplace, the car, or on the go.
For these user researchers, social distancing has reduced their ability to conduct effective user research.

With these challenges though, there is an opportunity to rethink how contextually situated user research is done.
I believe that there is a need to explore new directions in remote user research where researchers can conduct their work from anywhere with an internet connection.
These new research directions will build on prior work to develop new tools, methods, and frameworks that allow user researchers to conduct contextually situated work without needing to be on location.
This research will build knowledge about the strengths, limitations, and strategies for conducting effective remote user research that maintains the validity of research results and supports the development of new human-centered products and services.
Finally, this research will provide an opportunity for us to explore how the user research profession can be more sustainable and accessible.

In this position statement, I outline prior work in remote qualitative user research methods and discuss how this work can help design teams rethink their practices today under social distancing.
Specifically, I look at well-established methods such as experience sampling and remote contextual inquiry and discuss what previous research has found in regards to their effectiveness.
I then look at more recent developments in remote user research methods that employ new technologies to support user researchers in conducting research in complex environments.
The review of prior work then suggests a set of research questions that can further our knowledge and abilities to conduct remote user research.
These research questions include explorations of new systems that allow for better contextual understanding, studies on the effectiveness and limitations of remote methods, and research into how remote user research methods can be (re)designed to be more accessible and sustainable.
With new technology products increasingly augmenting our daily lives, good user research will be more important than ever.
However, doing great user research may not always mean we must be co-located with our users.

\section{Relevant Body of Work}
\label{sec:realted-work}

\textit{User research} is research done to understand people's experiences, needs, and goals in order to inform the design of products and services \cite{kuniavsky_observing_2003,Schumacher2009global}.
To conduct their work, user researchers employ a variety of qualitative and quantitative research methods ranging from early-stage needfinding \cite{patnaik_needfinding:_1999}, focused observation and feedback sessions \cite{beyer_contextual_1997}, lab-based usability studies \cite{nielsen1994usability}, and statistical analyses of behavioral data \cite{sauro2016quantifying}.
While a rigorous user research process employs both qualitative and quantitative methods, many practicing user researchers specialize in one or the other.
This position statement focuses primarily on the work that \textit{qualitative} user researchers do as it is more often done directly with users in-context rather than through analyses of user data and behavioral logs.
This focus on in-context work means that qualitative research methods are more likely to be impacted by social distancing and remote research operations due to it being harder to  travel and be co-located with users.


\subsection{User Research Methods for Contextual Understanding}
\label{sec:user-research-methods}
Much of qualitative user research in product and service design is inspired by ethnographic field methods that focus on capturing a holistic and descriptive view of people's lives in their natural settings \cite{blomberg1993ethnographic}.
Ethnography often takes the form of longitudinal, field-based observations and interactions with people to understand their everyday experiences and their point-of-view \cite{dourish_implications_2006}.
Within professional design teams, similar goals are met through ethnography-inspired qualitative field-studies in the form of ``rapid ethnography,'' with field-based work often lasting a few days \cite{millen_rapid_2000,plowman_ethnography_2003} rather than the more extended deployments of academic ethnography.
These rapid qualitative methods have become core research activities for professional design teams \cite{laurel_design_2003,kuniavsky_observing_2003,kelley_art_2007,sas_generating_2014}
as they allow a team to quickly understand the relationship that people have with products and services in the larger context of their lives \cite{beyer_contextual_1997, dourish_2004_action} and help the team develop empathy for their users, a core tenant of modern design thinking \cite{patnaik_needfinding:_1999, kelley_art_2007}.
Furthermore, field-based qualitative studies allow teams to understand the use of new products outside of the lab, helping them find social and contextual issues and interactions that are nigh impossible to see without being in the complexities of the real-world \cite{carter_exiting_2008,hornecker2012Lab}.

At their most basic level, most qualitative user research methods champion going out to the field to observe and interact with people directly.
These activities include interactions with people in their community and increasingly means traveling internationally to engage with users on the ground.
However, with almost all travel stopped and with social distancing measures limiting even local interactions, qualitative user researchers are forced to either halt their research operations or adapt them to remote work practices.
Fortunately, widely available internet and computer ownership means that many user research teams can move their operations online.
This is especially true for teams working on digital products and services but also includes teams working on physical products and services as well.
There is much research done over the years to develop and study remote research tools that enable design teams to conduct qualitative user research without needing to be co-located with the user.

\subsection{Established remote qualitative user research methods}
To overcome the challenges of being in the field with users, researchers and practitioners have developed remote user research methods and systems to capture qualitative experience data \cite{black_remote_2017}.
These methods can be \textit{asynchronous}, where users capture their experience without interacting with the researcher at the same time, or \textit{synchronous}, where researchers interact with users in real-time to understand their experience.
In the following sections, I describe some of the more commonly used remote qualitative user research methods.

\subsubsection{Experience sampling}
One of the earliest forms of conducting asynchronous remote qualitative research to understand people's everyday experiences is the \textit{experience sampling method}, developed by \citeauthor{larson1983experience} \citeyear{larson1983experience}, where research participants are periodically asked to document their experience and answer a set of questions when they receive a notification on a personal communication device, such a pager.
Experience sampling was further developed by \citeauthor{consolvo_using_2003} \citeyear{consolvo_using_2003} for use on mobile phones.
By using people's mobile device, researchers can capture text-based descriptions or conduct quick surveys of people's experience in the moment.
With the proliferation of mobile phones and open source software, experience sampling has become nearly free to implement with a wide range of users.
However, text-only interaction limits the view of the places and spaces that a remote researcher can experience and requires participants to describe their experience in more detail to remote researchers.
\textit{MyExperience} by \citeauthor{froehlich_myexperience:_2007} \citeyear{froehlich_myexperience:_2007} and \textit{Momento} by \citeauthor{carter_momento:_2007} \citeyear{carter_momento:_2007} expand upon these text-based experience sampling  by leveraging the added capabilities of mobile devices with cameras, sensors, and higher speed data to allow users to capture images, video, and data logs of their experiences.
Today, there are a number of commercial and open source apps that researchers can use to conduct multi-modal experience sampling such as \href{http://www.experiencesampler.com}{Experience Sampler}, \href{https://www.lifedatacorp.com/experience-sampling-app-2/}{LifeData}, and \href{https://ilumivu.com/solutions/ecological-momentary-assessment-app/}{mEMA}.
Experience sampling has the benefit of capturing user experience in the moment without the researcher needed to facilitate the interaction.
However, even with the added benefit of video and sensor logging, experience sampling can be limited due to randomly sampling and missing critical movements of a user's experience, users being preoccupied in another task and ignoring an experience sample, and the researcher not being able to follow-up on questions that they may ask given the answers that are received.
Due to these limitation, user researchers may look to supplement their research through more synchronous methods.

\subsubsection{Remote contextual inquiry and usability studies}
Remote contextual inquiry \cite{english_remote_2004} and usability studies can supplement the limitations of experience sampling by having the researcher interact synchronously with a user, allowing them to observe a user's experience and interview people in real-time.
Within software product development, researchers have explored how remote screen sharing and video conferencing could be used to develop desktop applications \cite{hartson1996remote,english_remote_2004} and mobile websites \cite{waterson_lab_2002,burzacca_remote_2013}.
Overall, these methods are now common practice among software development teams and are supported by many purpose-built user research tools such as \href{https://www.usertesting.com/}{User Testing}, \href{https://www.userzoom.com/user-research-methods/}{UserZoom}, and \href{https://lookback.io/}{Lookback}.
Furthermore, these remote methods have enabled software development teams to conduct international user research \cite{dray_remote_2004}, helping them reach a broader cross-section of their users while reducing the need to travel.
Under the current situation of social distancing, web-based video conferencing and screen sharing is the most readily available replacement for in-person research.
However, many teams may ask how effective research done through video conferencing and screen sharing may actually be?

\subsubsection{Effectiveness of remote user research}
When testing the software \textit{usability}, prior research has shown that \textit{synchronous} remote methods, including talk-aloud protocols and screen sharing, can be as effective as in-person studies in finding usability issues \cite{brush_comparison_2004,andreasen_what_2007,chalil2011synchronous}.
Interestingly, \citeauthor{brush_comparison_2004} \citeyear{brush_comparison_2004} report that most participants in their comparative study of remote and in-person methods preferred the remote study over the local study. 
Participants felt little difference in their ability to complete the study and felt less of the researcher looking over their shoulder, suggesting that remote studies can provide a better experience while also allowing the researcher to gain more access to the user's real-world environment.
Participating remotely can also be better for users who might otherwise not be able to come to a research team's location \cite{andreasen_what_2007}.

Despite these benefits, conducting remote studies can be more work for researchers and participants and can surface fewer usability issues \cite{andreasen_what_2007,bruun2009let}.
Additionally, much of the prior work on remote user research has focused on quantitative aspects of usability rather than qualitative factors \cite{sauer2019extra}, making these methods potentially less fruitful for early-stage formative research and for understanding more complex aspects of user experience and user's lives.
For example, while conducting remote studies with disabled participants, \cite{petrie2006remote} found that remote methods were less effective at providing rich qualitative data than in-person methods due to limitations in seeing the complexity of the user's real-world contexts.
Many of the aspects that make video-conferencing, screen sharing, and data logging so readily available for research teams such as using the cameras built into a computer or phone and studying software on the same device can limit researcher's ability to capture a broader picture of the user's environment.
These limitations have inspired other research into more experimental methods and technological systems for interacting with users remotely.

\subsection{Emerging Remote Qualitative Research Methods for Complex Environments}
Researchers have explored the use of new technologies and strategies for using communications systems to overcome the limitations of more standard experience sampling methods and video conference based remote user observations.
In the space of asynchronous remote user research, new advancements in chatbot technology are enabling more complex conversational interaction with with users, extending the simpler experience sampling methods of capturing answers to fixed questions.
Researchers are also adapting synchronous remote user research methods with extending Wizard-of-Oz practices where researchers play the role of and remotely control new technology to elicit interactions with and observe users acting in the real-world.
Finally, with the increasing amount of computation and sensing available in digital products, researchers are also co-opting these technologies to turn these things into co-ethnographers that can live in a user's environment and autonomously conduct field research, capturing realistic data and reporting back to the research team.
Overall, these methods each leverage new technological capabilities to extend the reach of a user research team.

\subsubsection{Chatbot assisted user research}
Text-based chatbots and vocal speech agents are being used to extend experience sampling methods and allow for more conversational interaction bewtween users and the research team.
These methods allow for experience sampling to move beyond fixed questions asking towards giving research teams the ability to ask follow-up questions and steer conversation based on the user's responses.
In an early example of a chatbot based user research interaction, \citeauthor{broadman_chatbots:_2016} \citeyear{broadman_chatbots:_2016} of IDEO used remote-controlled mobile phone-based chatbots to explore different services for health and wellness, having the remote team dynamically try new conversations to help determine user's needs.
Although remote controlled, the use of a chatbot as opposed to a fixed experience sample allowed the team to dynamically adjust their research questions over time and allowed them to try new sampling techniques.
In other work, \citeauthor{tallyn2018ethnobot} \citeyear{tallyn2018ethnobot} used a simple automated mobile phone-based chatbot, Ethnobot, to explore people's experiences at a festival, using their location to initiate questions.
Although simple, Ethnobot shows the possibility of leveraging the contextual data on mobile deices, such as location, to drive experience sample queries and to ask users questions at the moment they are in a specific context.

There are still open questions, however, as to how effective chatbot based qualitative methods are as it is quite challenging to mimic the thoughtful question asking behavior of a well-trained user researcher.
In the case of Ethnobot, while users interacted with the bot for an average of 120 minutes during the festival, many users were frustrated with the bot's rigid question asking and inability to build on the things that the users talked about in their responses \cite{tallyn2018ethnobot}.
Despite the reports of these challenges, other work has shown more advanced chatbots to be effective at capturing conversational and realistic information from users.
In studies of chatbot-based surveys, \citeauthor{kim2019chatbotsurvey} \citeyear{kim2019chatbotsurvey} found that chatbots with a casual conversational style collected higher-quality data.
More recently, \citeauthor{xiao2020tell} \citeyear{xiao2020tell} recently showed similar results in a comparative study of an AI-powered chatbot versus a typical web-based survey of open-ended questions, finding that users were more engaged and produced higher quality responses with the chatbot.
Although it is still early in our understanding of how well machines can hold qualitative conversations with users, these initial results suggest that ``a chatbot could perform part of a human interviewer's role by applying effective communication strategies.'' \cite[pg. 1]{kim2019chatbotsurvey}.

\subsubsection{Wizard-of-Oz in the wild}
In parallel to the development of chatbot based user research methods, ubiquitous computing and augmented reality researchers have built more interactive remote user research systems by making use of Wizard-of-Oz methods \cite{dahlback_wizard_1993} in real-world environments. 
For example, early work by \citeauthor{dow_exploring_2005} \citeyear{dow_exploring_2005} observed users and prototyped an audio-based spatial narrative experience for visitors to a local cemetery.
While this early work was done with a researcher following out-of-sight of the user, it laid the groundwork for remote Wizard-of-Oz research conducted from anywhere with a network connection.

The availability of high-speed cellular networks and ever-improving computing, sensing, and video capabilities of devices in our physical world now allow researchers to conduct remote research with multiple video, audio, and data streams from a user's environment.
More recent work has explored using multi-modal systems to enable remote user researchers to conduct contextual inquiry and interaction prototyping in complex real-world environments.
One such system is WoZ Way \cite{martelaro_woz_2017}, which leveraged real-time video and data transmissions to understand the user experience of cars with advanced driving assistance features and to explore music listening experience in the car and the home \cite{martelaro2020music}.
Through the use of multiple video, audio, and data channels all streamed live to a remote researcher's workstation, WoZ Way allows remote researchers to conduct rich qualitative research with users in their own environments.
Furthermore, the team also reported that they system allowed them to collaboratively conduct design research while being physically distributed in three different locations.

\subsubsection{Things as co-ethnographers}
With software and hardware products gaining more sensors and increasing levels of AI, researchers are also redefining the role things have during qualitative user research.
Computationally enabled products are now becoming part of the research process and not just artifacts that are studied during the research process.
\citeauthor{giaccardi2016things} \citeyear{giaccardi2016things} explore and call this idea ``things as co-ethnographers,'' where products imbued with sensors and cameras collect information in people's everyday environments such as the home.
Using things as co-ethnographers, research on everyday life is viewed from the perspective of augmented products. 
It gives the remote research team longitudinal access to user experience and provides ``unique insights about the relationships between objects and human practices'' \cite[pg. 245]{giaccardi2016things}.
\citeauthor{gorkovenko2019supporting} \citeyear{gorkovenko2019supporting} extend the use of thing co-ethnographers to provide remote user researchers with real-time data.
Their team used a speaker embedded with motion sensors to stream data to a research dashboard, allowing the research team to view the current music and the interactions that users had with the device.
The remote researchers then used data from the speaker to facilitate remote contextual inquiry through instant messaging, allowing users to respond to the researcher's questions through text, photo, or video.

The Wizard-of-Oz work above also has smart devices assume co-ethnographer roles and blends it with the method of using chatbots as user research agents by having the things ask user's questions about their experiences.
During \citeauthor{martelaro_woz_2017} studies on understanding driver experience and interaction with music streaming service, the research team's needfinding conversations were remotely mediated through the products themselves.
This moved the product from merely monitoring users to thing co-ethnographers as conversational partners \cite{martelaro_needfinding_2019}.
Recent advancements in chatbots-based interaction suggest a future where thing co-ethnographers can take a more active role in learning about and documenting in-situ user experience.

Overall, these trends in conducting remote user research with smart devices, remote data capture, and real-time communication are forming what \citeauthor{murray-rust2019entangled} \citeyear{murray-rust2019entangled} call \textit{entangled ethnography}.
This vision of entangled ethnography looks to develop research practices ``where constellations of people, artefacts, algorithms and data come together collectively to make sense of the relations between people and objects.'' \cite[pg. 1]{murray-rust2019entangled}.
The use of things as co-ethnographers and an entangled ethnography framework may help remote user researchers design research practices that better use the devices in people's environments and allow for them to conduct richer qualitative research remotely.

\subsection{Pros and Cons of Different Remote User Research Methods}
While each of the methods discussed allows for researchers to collect some form of in-context qualitative data about their users, each have their limitations. 
Table \ref{tab:procon} shows a basic list of pros and cons for each method.
Teams should consider their specific context and what kind of data they are looking to collect when choosing a method to implement.
For example, if the team has specific questions they know they want to answer, experience sampling or chatbot assistants might be a good choice.
If a team is looking to do open-ended research, remote contextual inquiry or things as co-ethnographers may be better.
Furthermore, the length of time to observe and interact with users is also important.
Remote contextual inquiry and remote Wizard-of-Oz may be good for short engagements whereas experience sampling or things as co-ethnographers may be more appropriate for longitudinal studies.
Finally, the team should consider how much technology they would like to build and employ.
Experience sampling and remote contextual inquiry are well-supported with readily available tools that can be implemented in an organization quickly. 
But, researchers may reach their limits sooner when limited to fixed questions or fixed views of a user's environment.
Remote Wizard-of-Oz, chatbot assistants, and things as co-ethnographers may require more custom hardware and software as well as a robust communications system.
Although more technically involved, these methods may provide researchers with a richer view of user's environments.

\begin{table}[t]
\caption{Pros and Cons of Different Remote User Research Methods}
\label{tab:procon}
\begin{tabular}{@{}lll@{}}
\toprule
\textbf{Method}                & \textbf{Pros} & \textbf{Cons} \\ \midrule
Experience Sampling Method     &  Asynchronous data collection & Missed moments \\
                               & Longitudinal data capture &  Takes user away from activity \\
                               &  Many tools available today  &  No in-the-moment follow-up \\ \midrule
Remote Contextual Inquiry      &  Live view of experience  &  Hard to do outside software \\
                               &  Can ask user to elaborate  &  Tied to a computer or phone \\
                               &  Reduce observer effect  &  More work for participants \\ \midrule
Remote Wizard-of-Oz            &  Live interaction prototyping &   Complex communication setup   \\
                               &  Test interaction in-context &  Challenging for researchers    \\
                               &  Explore complex environments &  Lost comm. breaks interaction   \\ \midrule
Chatbot Assisted User Research &  Scaled data collection &   Fixed questions   \\
                               &  More engaging than survey &  Needs more research on validity    \\
                               &  Good for specific questions &  Less observational   \\ \midrule
Things as co-ethnographers     &  Longitudinal data capture &   More work needed on privacy   \\
                               &  Access to more environments &  Limited field-of-view    \\
                               &  Collect quantitative usage data &  Complex to build device \\ \bottomrule
\end{tabular}
\end{table}

\section{Current Implications}

\begin{figure}
  \begin{center}
    \includegraphics[width=0.50\textwidth]{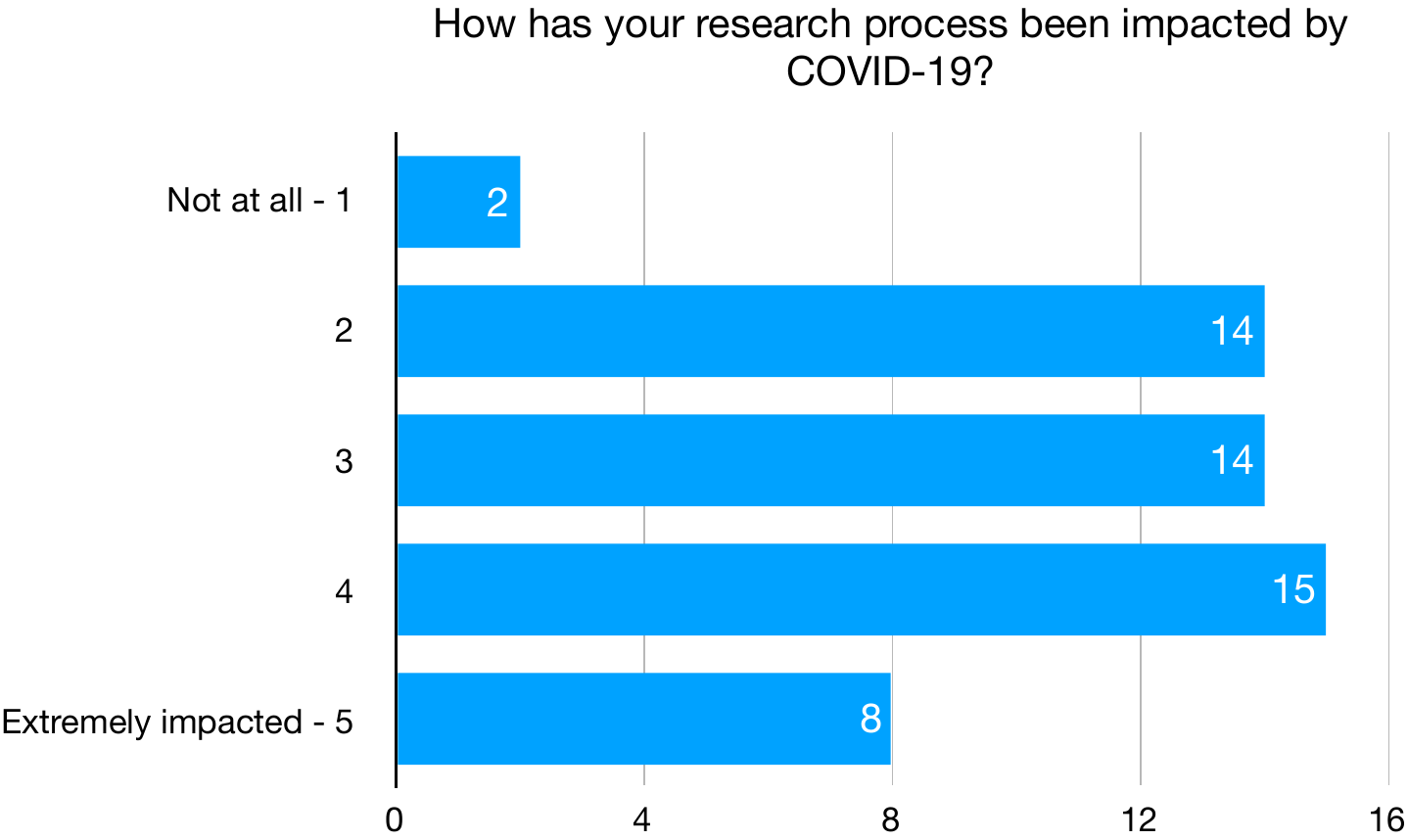}
  \end{center}
  \caption{Survey responses (N=53) on how severely COVID-19 has impacted user researchers work practices.}
\end{figure}

The COVID-19 pandemic and social distancing restrictions have significantly impacted qualitative user researcher's ability to do in-person fieldwork.
Research teams are not going into people's homes, riding along with people in their cars, or getting on-location in different cities or countries.
In my conversations with user researchers after social distancing policies were enacted, research teams were quickly reworking their plans to enable remote work.
Some teams, especially those who work on software, already have remote research operations using video conferencing and screen sharing in place and are further leveraging these capabilities.
However, other research teams who work with hardware or in environments such as the car are limiting their work to remote interviews and surveys or are suspending research operations temporarily.

We recently conducted an online survey \footnote{Responses collected between May 07 2020 - Jun 15 2020} collecting 53 user research professionals (primarily from the US) responses to questions on the transition to remote work.
When asked ``On a scale of 1 to 5 (1 = not impacted at all; 5 = extremely impacted), how has your research process been impacted by COVID-19?'' 23 researchers responded their work was \textit{heavily impacted} (15 responses of (4) and 8 responses of (5)), 28 researchers responded their work was \textit{somewhat impacted} (14 responded of (2) and 14 responses of (3)) and only 2 researchers said their work was \textit{not impacted at all}.
Respondents discussed canceling plans to travel to different countries, having difficulty recruiting participants, conducting ineffective remote contextual inquiry sessions over video chat, and finding it hard to collaborate and share qualitative findings with their other team members.
While the current situation is clearly a challenge for many teams, less impacted researchers often reported that they were already using remote research tools such as UserZoom and UserTesting.
Those who did not have these tools in place are now learning about and working to implement remote research tools.
Our survey found that 5 people were ``learning new tools and activities for remote research'' right \textit{before} social distancing policies were put in place while 25 researchers were learning new tools right \textit{after} social distancing policies were put in place.

Today, many teams are implementing more well-formed methods such as remote contextual inquiry on the desktop and experience sampling through mobile devices.
However, many researchers still find that these methods lack the qualitative richness that they would prefer when doing their work.
While user research teams are still in the early stages of figuring out their new normal, some that I spoke with have suggested that this is an opportunity for the user research profession to reconsider and (re)design their research practices for a world that may increasingly require and desire remote interaction with users.
As research teams continue to work remotely, we may see more examples of emerging methods in professional work.
Of course, rethinking our research process for remote work raises many questions for the field: Does qualitative work need to be done in person? Do we need to move so fast while conducting our research? How do we collect rich qualitative experiences and understanding when we can not be in the field?
How do we ensure we are collecting valid data about real user experiences?
Many of these questions will need to be answered to see a more robust framework for remote user research to take hold in industry.
For example, the present situation may accelerate the testing and acceptance of remote observation in complex environments and the increased use of things as co-ethnographers, however there is work to be done in understanding how these emerging methods work in professional design practice.

The current moment and the coming changes to remote user research processes allows us an opportunity to think about the desired future of the user research profession in regards to two other important considerations that will likely be influenced by the move to remote work: environmental impact and accessibility.
By moving research operations to remote practices, we can allow design teams to conduct work with global user groups without requiring extended international travel.
Reducing travel has the potential to reduce carbon emissions and can be a first step in helping teams to create sustainable design processes.
Furthermore, by enabling more remote work, we can also make the user research profession open to more people who may have otherwise found fieldwork challenging.
During the COVID-19 pandemic, many jobs that may have been considered too hard to be done remotely, and thus hard for those who are disabled, have now made many accommodations for remote work \cite{ishisaka_coronavirus_2020}.
We can take this opportunity to design new remote research practices that are more inclusive for user researchers with different abilities.
For example, researchers who use wheelchairs may have found it hard to travel to user's locations or access user's environments before but can now conduct remote contextual inquiry in a user's environment from their own locations.
Alternatively, researchers who may have trouble communicating directly with people due to limited sensory abilities could use smart devices, digital sensors, and chatbots to help them better engage with their users.

\subsection{Call for research and development}
I believe that the current moment is an opportunity to research and develop new tools, methods, and frameworks for doing remote user research.
The literature on ethnographic and user research methods \cite{blomberg1993ethnographic} offers a guide for field-based qualitative work as being:
\begin{enumerate}
    \item \textbf{Descriptive:} Researchers can see and record user behavior and attitudes
    \item \textbf{Holistic:} Researchers can understand user behaviors and attitudes in relation to the larger context and systems in the user's life
    \item \textbf{Done in the user's natural environment:} Researchers are embedded in the user's natural environment so that they can experience the user's life more closely to the user's perspective
\end{enumerate}
While being co-located with users is the most straightforward way to achieve these goals, it is not a strict requirement.
Instead, these principles can guide us in future development and studies of remote user research.

The various remote user research methods described earlier attempt to provide researchers with the ability to conduct descriptive, holistic work done in the user's environment even if they cannot be there in person.
While video conferencing and screen sharing are a start, some of the researchers I have spoken with and who responded to our survey found these environments to be too limiting. 
They expressed a desire to be able to see more of the user's context.
To enable more holistic user research, we will need to disconnect from desktop and mobile interfaces and capture a broader picture of our users' environments. New tools should allow us to follow users where they want to be rather than limiting their interactions with the remote research team to their computer or phone.
The prior work on experience sampling, remote contextual inquiry, and remote wizard-of-oz methods suggests ways to capture and understand a user's context when we augment user's environments with photo, video, audio, and data capture capabilities.

Going further, emerging ways of conducting user research through AI-enabled chatbots and smart devices can further allow remote research to be done in more complex environments.
By incorporating things as co-ethnographers, research teams may be able to access more complex environments, conduct research over longer deployments, and interact during more convenient and important times for their users.
Early work suggests that these methods may even support better qualitative data capture and description than when researchers are in-person or conducting survey-based research.

\subsubsection{Open research questions in remote user research}
There are several open questions that should be addressed as we (re)design our user research process to enable better remote work.
Below, I list what I believe are some initial guiding questions for research and development in the space of future remote user research methods.

\begin{itemize}
    \item \textbf{What information do in-person researchers attend to and get the most value from? How can we capture and transmit this information in remote settings? \cite{petrie2006remote}}
    User researchers often rely on non-verbal cues, broadly observing the user's context, and experiencing the physical sensations of a user's environment to develop a full sense of user experience.
    By understanding what researchers attend to and value, we can consider ways to design remote tools that provide these experiences directly or through proxies.
    Also, by collaboratively building new methods with professionals, we can leverage their intuitive sense of what is important and valuable to add these capabilities to remote research tools.

    \item \textbf{How effective are different forms of emerging remote user research methods in capturing valid data compared to in-person methods?}
    As the ecosystem of remote user research methods grows, choosing appropriate methods will require a user research team to understand what kinds of data they can collect and how valid their findings are given a particular context and research method.
    Currently, there are few comparative studies or measures of the effectiveness of different qualitative methods in practice.
    This is especially true for emerging methods such as using AI-based chatbots and conversational smart devices as co-ethnographers in complex open-world environments.
    While prior work has shown the opportunities of these new methods, more studies on their effectiveness and limitations will help professionals incorporate them into their user research activities.
    
    \item \textbf{What opportunities do remote methods afford that are improvements over in-person methods?}
    The prior work developing new remote user research methods have shown some capacities that are often not possible using in-person methods, such as capturing aspects of user's lives during all hours of the day \cite{giaccardi2016things}, collaboratively working with distributed research teams \cite{martelaro2020music}, and reducing the observer bias on participants \cite{brush_comparison_2004}.
    As more researchers move to remote methods, there is an opportunity to document what other benefits there are for teams such as cost, time, access, data richness, and ease for participants.
    
    \item \textbf{What tools will allow qualitative researchers to build and deploy their own remote research systems?}
    Remote research methods are all mediated by technology.
    This means that user researchers must either find and learn how to use tools that fit their intended workflows or build their own tools to meet the needs of their specific user context.
    There are opportunities for others to develop extensible tools and frameworks for allowing people to build remote research systems, even if the research team has limited technical capabilities.
    
    \item \textbf{How do different remote user research methods compare in regards to the user's experience while participating in the research?}
    As we use and develop our remote research practices, we must also take into consideration the user's experience as a participant.
    In many cases, remote research may be more convenient and less intrusive than in-person studies. 
    However, remote research methods may also be more time-consuming, cumbersome, or intrusive, depending on how they are executed.
    Studies evaluating new methods should look to capture the user's experience as a participant and provide guidance on picking appropriate methods to use in practice.
    
    \item \textbf{How can we conduct data-enabled remote research ethically? \cite{murray-rust2019entangled}}
    Remote user research methods inherently require data capture and transmission.
    As such, we should take care to respect user privacy.
    Future work should explore how we obtain continuous consent from participants and help participants understand how their data is being collected.
    Moreover, we should develop methods that foster conversation and interaction with users rather than surveillance and monitoring of users.
    
    \item \textbf{What are the environmental impacts of our current user research processes and how can we reduce our impact through remote methods?}
    Although we know that all travel contributes to carbon emissions, few studies document user research's environmental impact.
    For small teams with infrequent travel, the impact may be low, but larger organizations with global user bases may have a higher impact.
    Remote research operations have the potential to reduce carbon emissions significantly.
    Future research should look to estimate the impacts of current in-person research and how environmental impacts could be reduced with remote research practices.
    
    \item \textbf{How should remote user research activities be designed so that they are inclusive and available for user researchers who might otherwise find fieldwork challenging?}
    As with many other forms of work, much of what we thought needed to be done in person can be accomplished remotely with proper accommodations.
    Future research should consider exploring who might currently be left out of field-based user research teams.
    We should then collaboratively redesign remote user research tools and methods with those of different abilities so that a more diverse set of people can work as user research professionals.
    
\end{itemize}

\section{Conclusion}
The COVID-19 pandemic has presented many challenges to qualitative user research by shutting down many in-person research activities.
Although the virus is temporary, this event will most likely change how many organizations conduct their user research and will push them to use and develop remote qualitative research methods.
Even when we can return to conducting in-person research, having a robust remote research portfolio can act as insurance against future restrictions.
It could also allow for a more cost-effective, environmentally conscious, and inclusive way for organizations to understand their user's experiences.
To develop new methods, we can build upon a strong foundation of prior work on experience sampling, remote contextual inquiry, and remote wizard-of-oz methods.
We also have an exciting new set of research methods and frameworks employing the increasing amount of computation and smart devices in our world to augment our research teams with new data collection and user interaction techniques.
As our current circumstances accelerate the move to remote user research, I look forward to contributing toward and seeing new developments in building better research processes to help designers keep up with our rapidly changing world.
With creative development, we can build remote research methods that afford new ways of understanding user experience today and for a future where remote research is much more common.













\articlebodyend


\begingroup
\raggedright
\titleformat*{\section}{\bfseries\Large\centering\MakeUppercase}
\bibliographystyle{apacite}
\bibliography{martelaro.bib}
\endgroup








\end{document}